# Mesoscale Polymer Arrays: High Aspect Ratio Surface Structures and Their Digital Reconstruction

**Authors**: Demi E. Moed[a], Michael S. Dimitriyev[a,b], Benjamin R. Greenvall[a], Gregory M. Grason[a]*, Alfred J. Crosby[a]*

[a]Department of Polymer Science and Engineering, University of Massachusetts Amherst
Amherst, Massachusetts 01003, United States

[b]Department of Materialse Science and Engineering, Texas A&M University, College Station, Texas, 77840, United States

Corresponding Author's Email: acrosby@umass.edu (A.J. Crosby), grason@mail.pse.umass.edu (G.M. Grason)

**Abstract**

Inspired by adhesive bio-filamentous structure, such as bacterial pili, this work details the methods used to fabricate and characterize a surface-anchored array of thin, flexible and shape-responsive mesoscale polymer ribbons with a length-to-thickness aspect ratio of up to 100,000. The resulting structures exhibit geometrically complex and dynamic morphologies consistent with elastocapillary bending that experience an increase in curvature over hours of aging due to creep. We develop a computational image analysis framework to generate 3D reconstructions of these densely crowded geometries and extract quantitative descriptors to demonstrate morphological changes due to aging. We demonstrate the robustness of this quantitative method by characterizing the creep-induced change in an aging ribbon array's shape and develop a scaling relationship to describe the importance of ribbon thickness for shape and dynamical observations. These methods demonstrate an essential baseline to probe morphology-property relationships of mesoscale polymer ribbon arrays fabricated from a variety of materials in numerous environments. Through the introduction of perfluorodecalin droplets, we illustrate the potential of these ribbon arrays towards applications in adhesive, microrobotic, and biomedical devices.

**Introduction**

Nature commonly uses high aspect ratio structures to mediate surface interactions. For example, it is widely known that the fibrillar arrays of setae and spatulae on gecko toe pads offer structural advantages when climbing by allowing the geckos' toe pads to make intimate contact with complex surfaces.[1–5] Bacterial pili, or rod-like organelles that boast an aspect ratio on the order of 100, are another example.[6] As demonstrated in **Figure 1a**, the high aspect ratio of pili allow them to densely populate the surface of bacteria.[7] They are one of the many tools that moderate bacterial surface attachment and adhesion.[6,8] Pili are tipped with adhesin proteins that bind to specific molecules on the surface of host cells.[8,9] The proteins that comprise these organelles are helically coiled, which allows the organelle to unfold under stress.[10,11] As a result of this plastic deformation, pili can elongate up to several times their original length before rupturing, which allows bacteria to maintain surface attachment in the presence of strong flow fields.[10,11]

The development of synthetic surface structures with similar aspect ratio, size scale, and large compliance opens opportunities for mediating surface interactions in a similar manner. For example, synthetic pili could be engineered to adhere at preferred attachment sites or respond to environmental changes.[12] Synthetic, ultra-high aspect ratio structures may also entangle with nearby materials to provide mechanical attachment mechanisms, such as wrapping, beyond simple van der Waals or elastic interactions.[13–16] Such capabilities are desirable for advanced robotic devices, adhesives, and capture-and-release systems.[12,13,17] However, the creation, characterization, and control of such structures offers challenges. It is difficult to moderate the impact of surface interactions between neighboring high-aspect ratio structures at these length scales, thus such interactions typically cause densely-anchored surface arrays to self-collapse or form complex contact arrangements.[17–19] However, the same sensitivity to surface interactions also presents an opportunity for dynamic control and modulation of collective large-scale morphology.[12,13,17,19–22]

Mesoscale polymer ribbons (MSP) offer a unique means of achieving the desired flexibility and high aspect ratio to create pili-inspired structures. MSPs have thickness ranging from ten to hundreds of nanometers, widths on the order of tens of microns, and lengths of millimeters or more.[23] The ratio of their length to their width approaches 100, and their width to their thickness approaches 1,000 (**Figure 1d**). This size scaling renders them susceptible to both macro- and microscale phenomena. They can be formed out of nanoparticles, quantum dots, or polymers, allowing access to a variety of mechanical, optical, and surface properties.[14,24–26] When submerged in liquid environments, the balance between their bending stiffness and surface tension results in the ribbons morphing into 3D geometries,[24,26] which can be tuned through materials selection and environmental conditions.[26,27] The elastocapillary-driven curvature of MSPs is an appealing parallel to the helical geometry of bacterial pili.

Here, we introduce MSP ribbon arrays, which are high aspect ratio polymer filaments that extend from one end anchored to a surface into a surrounding aqueous environment. The anchoring points are arranged in ordered arrays, but the length of MSP ribbons spans many times the spacing between neighboring attachment sites. This design, in combination with their extreme flexibility due to their high aspect ratio, allows them to offer promise as surface-active structures for mediating interfacial interactions, with potential impacts on adhesion, microrobotic, and biomedical devices.

We describe the methods of fabrication and characterization of MSP ribbon arrays and demonstrate their effectiveness through a case study of MSP ribbon array aging. The size and flexibility of these structures introduces new challenges for characterizing their structural morphology and how it responds to external and internal cues. We develop a new computational framework to use fluorescence confocal microscopy imaging to characterize and quantify how structural parameters, such as curvature, vertical mass distribution, and inter-ribbon separation, evolve. By introducing this framework, we not only learn how interfacial forces couple time- and size-dependent materials to define mesoscale structure but also introduce computational methods that can be transferred to filamentous or linked systems on other scales or contexts. We further demonstrate the potential functionality of surface-anchored MSP ribbon arrays by showing their ability to wrap and trap liquid oil droplets.

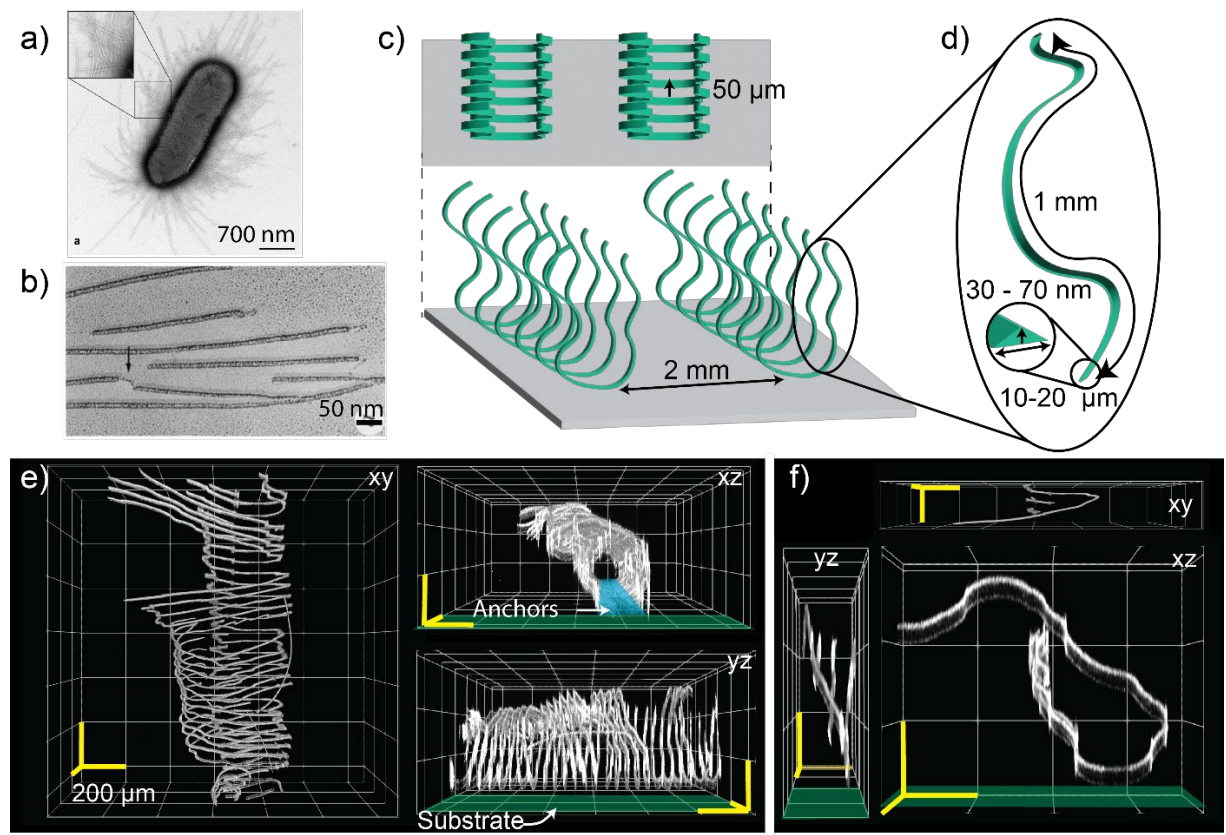

**Figure 1**. (a) TEM micrograph of *E. coli* and with type 1 pili. Reproduced with permission from [7]. (b) Electron micrograph of a plastically deformed *E. coli* P-pili. Reproduced with permission from [10]. (c) Poly(tert-butyl methacrylate) ribbons anchored in rows of 2 mm separation with ribbons spaced 50 μm apart. (d) A sample MSP ribbon with average lengths of 1 mm, widths ranging from 10 μm - 20 μm, and thicknesses ranging from 30 nm to 70 nm, and a characteristic triangular cross-section. (e) A 3D reconstruction of an MSP ribbon array, as imaged via confocal microscopy. The location of the substrate is indicated in green, and the anchors in light blue. (f). Confocal microscopy of a single ribbon from the array. All scale bars of confocal microscopy images are 200 μm.

**Experimental Methodology**

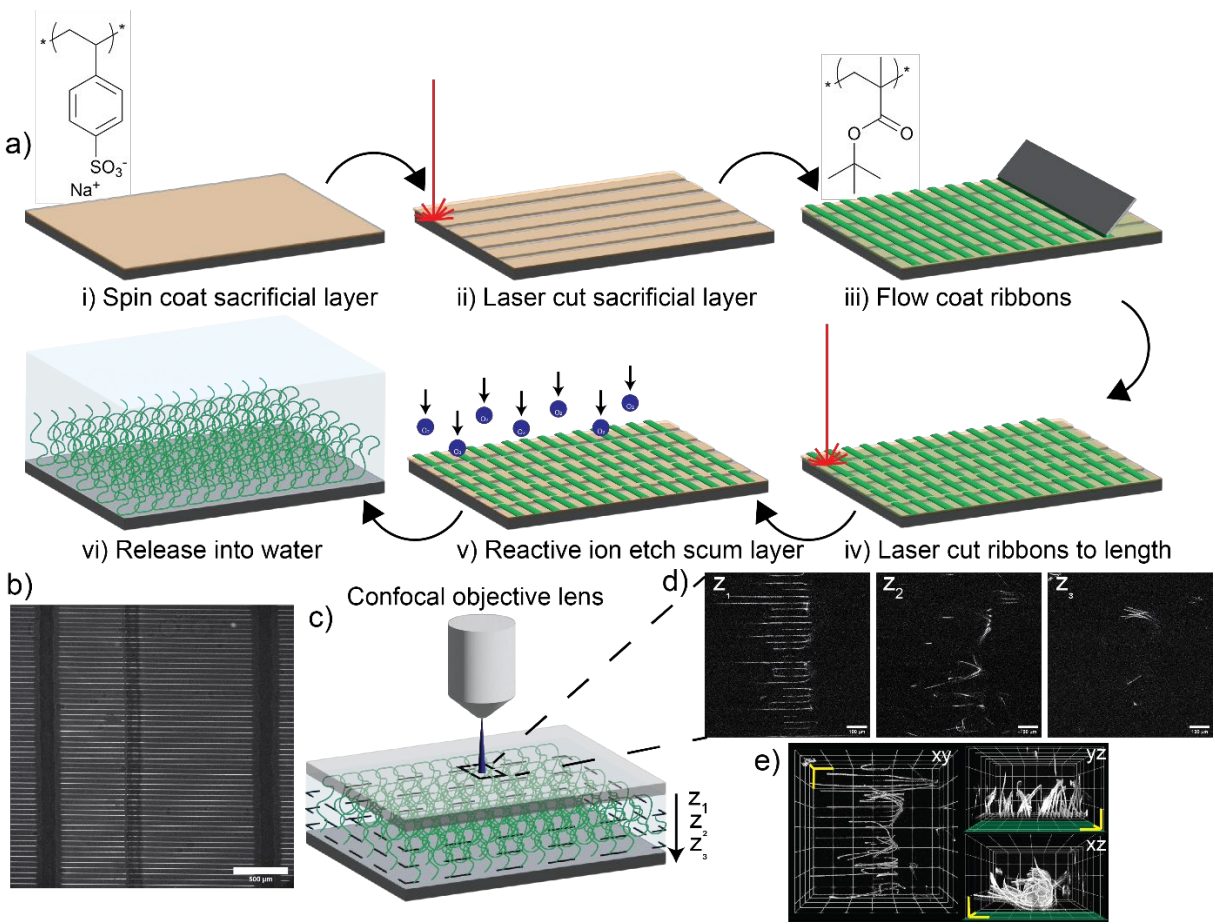

**Figure 2**. (a) An overview of the methods used to make mesoscale polymer ribbon arrays. (b) Prior to release in DI water, MSP ribbon arrays appear as a film of flat, horizontal lines broken by the vertical laser cutting lines in fluorescent microscopy. Here, the vertical line closest to the center corresponds to the ribbon anchor, and the two, darker lines to the far left and right create the ribbon tips. Scale bar: 500 μm (c) After release in water, MSP ribbon arrays are imaged via confocal microscopy. By scanning a region from top-to-bottom ($z$ positions $z_1$, $z_2$, and $z_3$) (d) individual slices (scale bar: 100 μm) are layered to create (e) a 3D reconstruction (scale bar: 200 μm) of the array.

*Substrate Preparation.* Glass cover slips (VWR #1.5 micro cover glass 24 mm x 60 mm x 0.17 mm) are cleaned via sonication for 15min each in soapy water, reverse-osmosis water, and isopropanol and then irradiated with UV-ozone (Jelight Company, Inc. Model 342 UVO-Cleaner) for 20 minutes to create a hydrophilic surface.[27] An aqueous solution of 20 mg/mL poly(sodium-styrene sulfonate) (Mw ~70kDa, Aldrich) is filtered through a 0.45 μm PVDF filter (Fisherbrand #09-720-4) and spin-coated onto the glass slides for 10 s at 500 rpm and then at 40 s at 2000 rpm. This creates a thin, water-soluble sacrificial layer (**Figure 2a.i**).

Using a laser cutter (Universal Laser Systems VLS3.50 with a 30W $CO_2$ 10.6 μm laser), straight lines spaced 2 mm apart are engraved into the sacrificial layer (**Figure 2a.ii**) at 3% power, 40% speed, and 1000 ppi. These engravings serve as anchoring sites for the mesoscale polymer ribbons. The substrates are then annealed on a hot plate at 130 °C for a minimum of 15 minutes to drive off excess water.

*MSP Ribbon Array Fabrication.* We use flow coating [23,27] to fabricate the mesoscale polymer ribbon arrays. During this process, a polymer solution is injected between an angled razor or silicon blade positioned just above the substrate's surface. A smarAct, Inc. SLC-1780s linear actuator then laterally translates the substrate in a stop-and-go motion at fixed distance intervals. Between movements, the polymer deposits along the three-phase contact line through evaporative assembly. The geometry and quality of the resulting MSP ribbon array is a balance between the solution concentration, step size, stop time, and step speed. The step size must be large enough to ensure distinct separation between neighboring ribbons. Furthermore, the stop time must allow for sufficient polymer deposition, as ribbons that are too thin will easily tear from the substrate during release, and the substrate must move fast enough to allow the capillary bridge between the blade and substrate to be ruptured to deposit independent ribbons.

We use a 2.3 mg/mL solution of poly(*tert*-butyl methacrylate) (PtBMA) (Mw ~ 170 kDa, $\rho$ =1.022 g/mL, Sigma-Aldrich) in toluene (Fisher Scientific) blended with trace amounts of coumarin-153 (Sigma-Aldrich) for fluorescent imaging. All arrays are deposited with an inter-ribbon spacing of 50 µm over a total distance of 5 mm. After each step, the actuator dwells for 2000 ms before laterally translating to the next location at a speed of 3 mm/s (**Figure 2a.iii**).

The ribbons and sacrificial layer are engraved for a second time under the same conditions. This subsequent engraving is performed parallel to the previous engraving lines and offset by 1 mm (**Figure 2a.iv**). This process cuts the ribbons to ~1 mm in length and creates a free end that is not anchored to the substrate. The slides are reactive ion etched (STS Vision 320 Mark II System) with oxygen plasma at 50 mTorr for 30 s at an RF setpoint of 100 W to remove the inter-ribbon scum layer (**Figure 2a.v**). We use optical profilometry (Zygo NewView 7300) to determine the dimensions of our ribbons prior to release (See **Figure S1**). As there can arise differences in ribbon width and thickness across a sample due to changes in solvent concentration between the start and end of the flow coating process, the thickness and width values we report for each sample represent an average of multiple regions across a substrate surface.

*Ribbon Release and Characterization.* Each ribbon array is placed into a flow chamber of 25 mm x 75 mm x 1 mm glass slides (Fisher Scientific) connected by Loctite Quick Set Epoxy to ensure unidirectional flow (**Figure S2**) using the methods outlined in Barber et *al.*[26,27] The ribbons are imaged using fluorescent optical microscopy (Zeiss Axio Observer 7) prior to release (**Figure 2b**) to determine ribbon length. Grease wells are made at either end of the flow chamber. The chamber is filled from one end with DI Water (Alfa Aesar), which dissolves the sacrificial layer and lifts the ribbons from the substrate. The chamber is sealed by sticking parafilm to the grease wells to slow evaporation and prevent spillage. For the purposes of time-dependent analyses described below, the time of release is considered $t$ = 0 s.

After release into DI water, the mesoscale polymer ribbon arrays adopt a 3D conformation as elastocapillary forces bend them away from the substrate (**Figure 1e**). Accurate resolution of the arrays along the z-axis requires a narrower depth of field than fluorescent optical microscopy can provide. Therefore, we use a Nikon FN1 stand with a A1HD(1024) Resonant Scanning Multi-Photon (RMP) Confocal microscope with $\lambda_{ex}$= 488 nm to visualize the 3D morphology of the ribbon arrays. As shown in **Figure 2c**, each region of interest (1.27 mm x 1.27 mm) is imaged repeatedly at fixed intervals (4.65 µm step size) along the z-axis. These individual slices are then compiled into a z-stack from which the 3D morphology can be rendered (**Figure 2d,e**) using NIS Elements 5 software. Representative Z-stacks of a 36 nm-thick (**Movie S1**) and a 73 nm-thick (**Movie S2**) ribbon array taken immediately after release can be found in the supplementary information. To study the evolution of ribbon array morphology as a function of aging, we collect these z-stacks at half-hour intervals for up to 6 hours from the start of data collection.

*Computationally-Extracted Morphology.* As demonstrated in **Figure S4**, mesoscale polymer ribbon arrays exhibit remarkably varied appearances, even amongst regions of equivalent thickness. Therefore, we turn to quantitative features of intra- and inter-ribbon geometry to provide a means of bringing the dynamical and structural behaviors of these systems into focus. In order to calculate these metrics, we transform the 3D grayscale image stacks that result from confocal microscopy into a set of ($x,y,z$) positions describing the path of each ribbon, from which quantitative descriptors can be calculated. This process is complicated by apparently entangled structures frequently formed by the ribbon arrays. A detailed description of this methodology has been included in the SI. We begin this process by thresholding (**Figure 3a, b**) and then skeletonizing (**Figure 3c**) the images using the open-source Fiji (ImageJ) software.[28] Using strategies previous studies [29–34], a MATLAB (R2022a) algorithm then sorts the voxels of the resulting skeleton into ends, backbones, and junctions based on the occupancy of their neighborhood. Neighboring backbones are connected to create segments terminated at either junctions or ends. To represent the directionality of each of these segments in 3D space, we consider each filled voxel within a segment to have equal mass. We subsequently compute the gyration tensor of each segment and define its directionality by its largest principal eigendirection. To correctly pair ribbon segments at crossings, we rely on the tendency of each ribbon backbone to maintain its orientation. We prioritize forming cross-junction connections between segments of parallel directionality (**Figure S3**). This is because mesoscale polymer ribbons have a measurable bending stiffness, and therefore parallel connections require the least energetic cost. We connect the segments across a junction whose principal eigenvectors exhibit the dot product furthest from zero (i.e. whose eigenvectors are closest to colinear). The resulting buildups are then evaluated to ensure that the ribbons follow singular, continuous, and unique paths, and ribbons that fail these criteria are broken apart and re-evaluated. **Figure 3d** a representative 3D reconstruction of quantitative microscopy data.

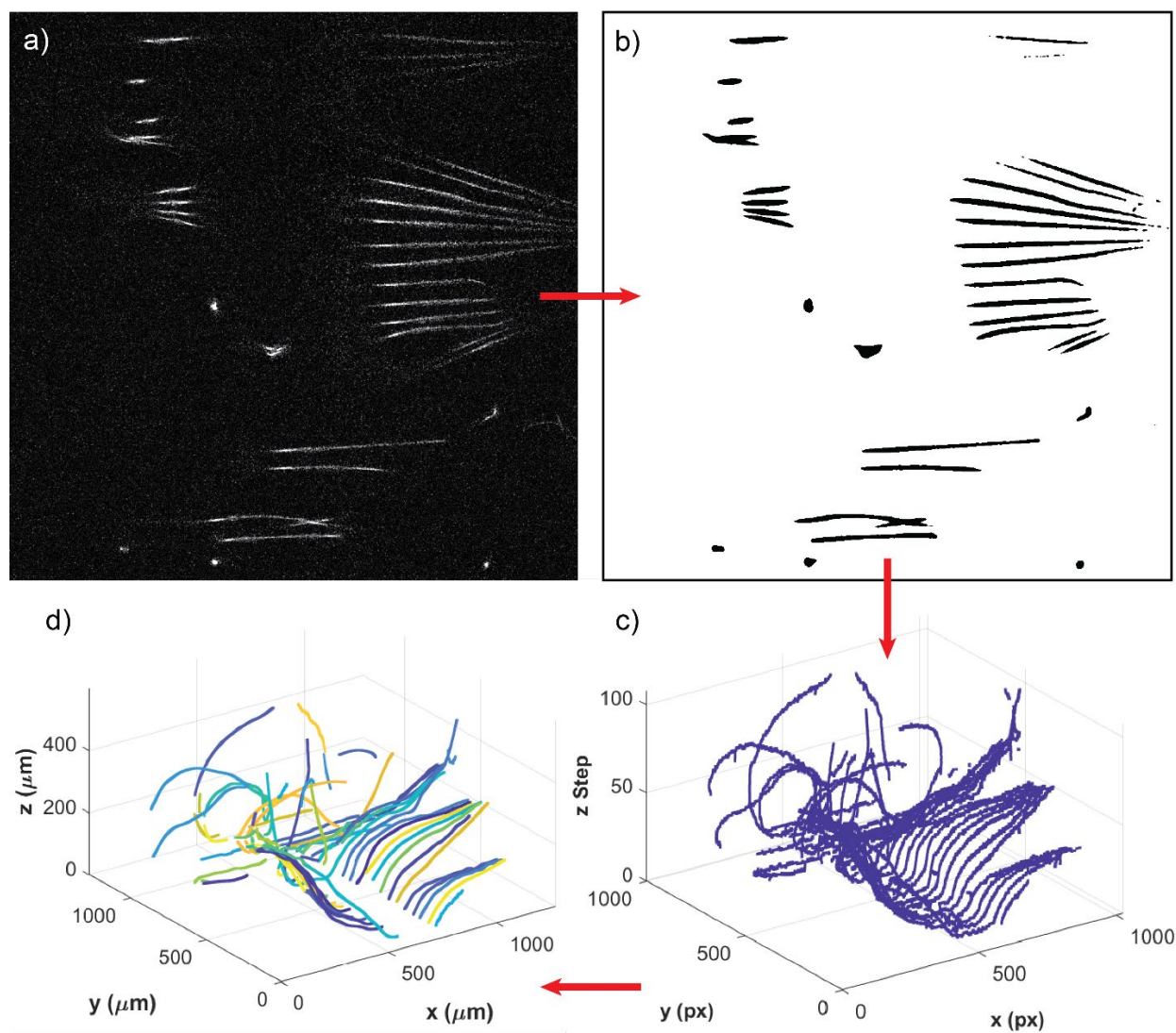

**Figure 3**. Each (a) raw data set is (b) thresholded prior to (c) skeletonization and processing, resulting in (d) a 3D reconstruction.

The algorithm then calculates key quantitative descriptors of ribbon array morphology. The average curvature of the mesoscale polymer ribbon array should relate to the balance between a ribbon's bending stiffness and surface tension with the ambient environment.[24,27,35,36] While we expect homogenous bending across the length of the ribbons, we find instead that ribbon lengths fall into two populations: locally curved and effectively straight. In these latter regions, the ribbon curvature approaches zero and the radius of curvature diverges. We expect that these regions correspond to imperfect dissolution of the sacrificial layer, and as such, they provide a means of measuring the success of the release process. We calculate this as the length fraction of each ribbon whose radius of curvature is greater than the target ribbon length. We also introduce two metrics with which to gauge the orientation of the MSP ribbon array in 3D space: the vertical mass distribution and the lateral mass distribution. For both calculations, we consider each individual voxel filled by a ribbon's skeleton to have an equal mass. To calculate the vertical mass distribution, we calculate the average vertical distance ($\Delta z$) between each point of mass and the substrate. To calculate the lateral mass distribution, we calculate the average lateral distance ($\Delta x$) between a ribbon and the central line of anchor points produced by the first laser cut. The end-to-end distance between a ribbon's anchor point and its tip provides further insight into the coiling of each ribbon. As a ribbon's end-to-end distance approaches its length, it adopts an increasingly rod-like morphology, whereas a comparatively small end-to-end distance relative to a ribbon's length indicates a coiled conformation. To measure inter-ribbon interactions, we calculate the average distance between each ribbon and its nearest neighbors, as well as the number of instances where two ribbons approach at distances small enough for surface interactions to occur. Finally, the average ribbon length of a quantitative reconstruction should provide a means to gauge the accuracy of the buildup, as it can be directly compared to pre-release microscopy measurements.

*Droplet Wrapping.* To provide evidence towards the potential application of mesoscale polymer ribbon arrays as filters and adhesives, we provide a proof-of-concept demonstration of their collective interactions with foreign objects. Arrays of mesoscale polymer ribbons were released into aqueous solutions of DI Water, pH 4 and pH 10 buffer (Supelco) by floating the glass substrate on the air-water interface and then rapidly submerging the slide using tweezers. Droplets of perfluorodecalin (Aldrich) were generated by rapidly agitating a mixture of perfluorodecalin and DI water. These droplets were introduced to the MSP ribbon array via Pasteur pipette. We used fluorescent optical microscopy (Zeis Axio Observer 7) to visualize the interactions between the mesoscale polymer ribbon arrays and the droplets of perfluorodecalin.

**Results and Discussion**

*Mesoscale Polymer Ribbon Array Fabrication*

We demonstrate the successful fabrication of substrate-bound MSP ribbons (**Figure 1e**). With a lateral inter-ribbon separation of 50 µm and an average ribbon length of 1 mm the ribbons detach from the substrate following the dissolution of the sacrificial layer in all regions not anchored to the substrate. The ribbons exhibited an average thickness between 30 nm and 70 nm and average widths ranging from 10 µm to 20 µm. The range in thicknesses and widths is a result of small differences in the flow coating process between samples. Although step size, stop time, and post-step delay are consistent across slides, slight deviations in solution concentration, ambient humidity and temperature, and the distance between the blade and the substrate during flow coating can result in differences in ribbon thickness and width between samples, and therefore are accounted for in subsequent analysis.[37]

We note following the release of the ribbon arrays into water that we occasionally observe that the ribbons are torn from the substrate with a point of fracture just beyond where they are tethered to the substrate (see **Movie S3**). We attribute this ribbon breakage to the drag and capillary forces that accompany the advancing aqueous meniscus. We observe that a rapid immersion process reduces this effect, and we postulate that this allows the 3-phase contact line to advance beyond a ribbon's length before the sacrificial layer fully dissolves.

*Ribbon Array Morphological Aging Dependence*

As seen in **Figures 4** and **S4**, the mesoscale polymer ribbon arrays undergo morphological changes as they age, with the most noteworthy change being an increase in ribbon curvature over the course of the 6-hour period. As the time it takes between ribbon release and the start of data collection varies between samples, all reported times are relative to the ribbon's immersion in DI Water. For videos of this process, please refer to **Movies S4**-**S7**.

To understand the origin of this phenomenon, we begin by turning to Pham et *al.*, who demonstrate that the curvature of a mesoscale polymer ribbon scales as:[24]

$$(1) \quad \kappa = \frac{\gamma P \Delta X_y}{E I_{yy}} \sim \frac{\gamma}{E h^2}$$

Where, $\kappa$ is the curvature, $\gamma$ is the surface tension, $P$ is the ribbon's cross-sectional perimeter, $\Delta X_y$ is the offset between the center of a ribbon's cross-sectional area and center of the ribbon's cross-sectional perimeter, $E$ is the Young's modulus, and $I_{yy} = \int dA \, y^2$ is the component of the second moment of area of the ribbon along its thickness direction.[24] By assuming that the width of the ribbon $w$ is significantly larger ribbon thickness $h$, the curvature equation is simplified to a function of modulus, thickness, and surface tension. This equation arises from the balance between a ribbon's desire to minimize its surface tension with the surrounding environment at the cost of inducing bending along its length. Thus, the curvature is defined by a balance between the magnitude of the ribbon's surface tension and its bending stiffness.

However, this formula does not account for the time-dependent increase in the curvature we observe in our poly(*tert*-butyl methacrylate) ribbon arrays. Previous work with poly(methyl methacrylate) (PMMA) MSP ribbons corroborates these results and provides an explanation.[35,38] Daïeff observes that the radius and pitch of MSP PMMA helices with thicknesses ranging from 15 nm to 35 nm evolved with time, with the radius achieving equilibrium on the order of hours and the pitch on the order of minutes.[38] After ruling out alternate explanations, such as sacrificial layer desorption and temporal variation of bending modulus, the time-dependent morphology change is attributed to creep. Prévost et *al.* subsequently leveraged creep to tune the pitch angle of mesoscale helices using uniform viscous flow.[35]

Unlike these previous studies,[24,35,38] our ribbons do not uniformly coil into helical structures, nor do they exhibit a uniform curvature across their length. Instead, upon release the ribbons bend into a variety of morphologies with inconsistent curvature along the length. These morphologies vary both across each region and between samples. For example, in **Figure 4**, several ribbons that are relatively flat against the substrate and parallel to other ribbons at earlier time stamps curl into relatively wide arcs at larger time stamps. For ribbons shown in **Figure S4a**, which share the same thickness as the ribbons in **Figure 4**, the ribbons near the top of the reconstruction exhibit a much smaller visual change and much higher overall curvature during aging than the ribbons towards the bottom of the construction. In **Figure S4e**, the mesoscale ribbons towards the bottom of the slide have already formed a tight coil upon the start of data collection, but the upper ribbons coil over a longer time. Ribbons in **Figures S4b** and **S4d** develop circular structures reminiscent of helices by the end of the aging experiment. To our knowledge, existing attempts to quantitatively model this creep dependence have been only somewhat successful.[38]

Some of these inconsistencies can be explained by slight variations in ribbon thickness across each slide. As shown in **Figure S1**, adjacent ribbons demonstrate subtle differences in their thickness, and individual ribbons can exhibit a range of thicknesses across their length. By Equation 1, these differences in ribbon cross-sectional geometry translate to changes in intrinsic ribbon curvature. Therefore, the curvature of our MSP's is a function of position along the ribbon backbone, with extreme values corresponding to the positions of minimum and maximum thickness. Furthermore, the flow chamber in **Figure S2** is not designed for uniform, laminar flow across the sample, and so differences in drag forces upon release may also account for differences in architecture, as strong drag forces may result in the plastic deformation of the ribbons. Owing to these variations between ribbons, we use the average curvature in subsequent quantification.

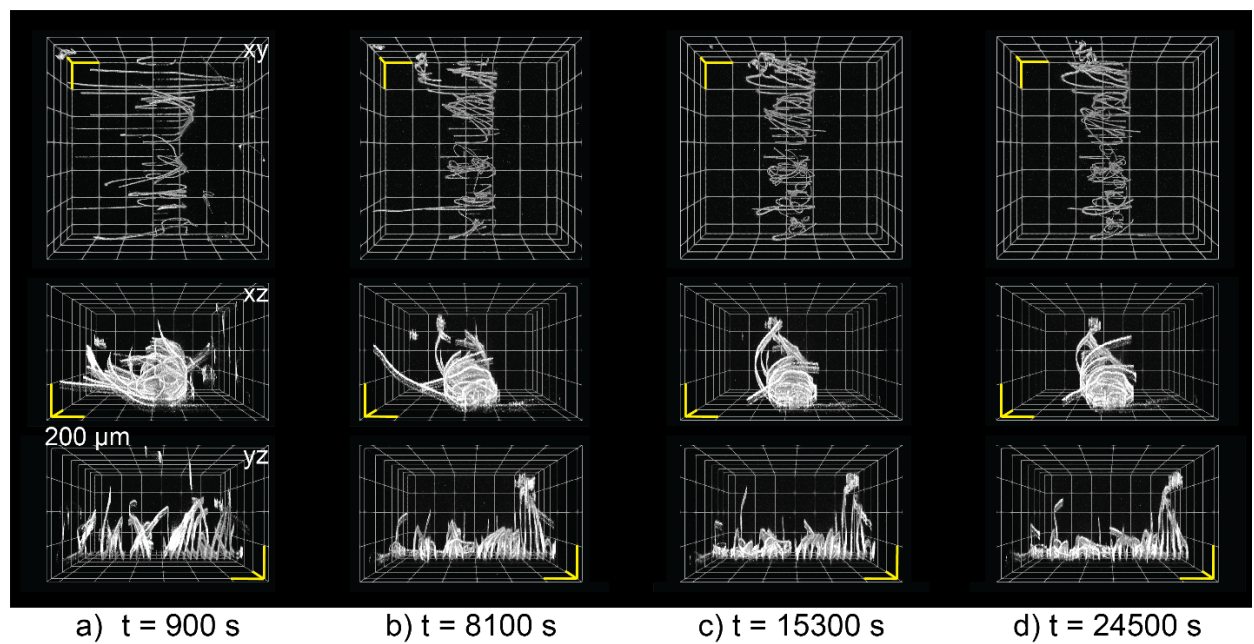

a) t = 900 s      b) t = 8100 s      c) t = 15300 s      d) t = 24500 s

**Figure 4**. Confocal microscopy image of an MSP ribbon array at a) 900 s, b) 8100 s, c) 15300 s, and d) 24500 s after release.

Confocal microscopy provides a robust means of visualizing the impact of aging on MSP ribbon arrays. However, as is demonstrated in **Movies S4-S7** and **Figure S4**, even under similar aging conditions, MSP ribbon arrays exhibit a wide variety of architectures. Although it is relatively easy to visually track the changes in individual ribbon orientations as a function of time, qualitative comparison between regions with different starting conditions is quite imprecise. Therefore, using computational image analysis, we develop quantitative descriptors to identify average morphological changes to the mesoscale polymer ribbon arrays.

*Comparison between Confocal and Quantitative Models*

Visual comparison between the computationally-driven ribbon array reconstructions and the confocal data (**Figure 5 a, b**) indicate that the reconstructions are reasonable estimates of ribbon array morphology. A key feature of the computational reconstruction is the exclusion of the substrate-bound section of each of the ribbons. As these regions are unable to undergo conformational changes due to their adsorption to the substrate, we exclude them from the analysis of shape change. When crafting the code, we struck a balance between computation time and accuracy.

Since we expect the contour length of ribbons to be preserved when they are not subject to strong flows, we use this metric to quantitatively gauged the accuracy of the array reconstruction. In **Figure 5c**, we observe little correlation between ribbon length and time step, which is to be expected because the ribbons should not change in length as they age. However, when comparing these values to the expected ribbon lengths determined by microscopy (see $t$ = 0s), we notice a calculated length roughly 30% of the expected value (260 $\pm$ 60 μm of 870 $\pm$ 30 μm). This expected ribbon length is measured using optical microscopy prior to releasing the ribbons and represents the distance between the edge of the ribbon anchor and the tip of the ribbon and excludes the portion of the ribbon attached to the substrate. **Figure 5f** depicts the distribution of ribbon lengths of a set of 73 nm-thick ribbon arrays between 1300 s and 2100 s after release, and it is evident that most of the model ribbons exhibit lengths much smaller than the contour length measured prior to release. While some of these shortened segments result from ribbon breakages during release, ribbons exiting the field-of-view, and thresholding cutoffs, many arise from errors in the reconstruction algorithm. A small portion of the expected length is lost to the removal of the substrate-bound portion of the ribbons, and we anticipate many of these fragments result from errors in skeletonization and reconstruction parameters. For example, the hook-like ribbon in the middle-left hand side of **Figure 5a** is clearly a single, continuous ribbon, and yet in **Figure 5b** it is broken into multiple, smaller segments. Therefore, quantitative descriptors that are dependent on ribbon length, such as end-to-end distance (**Figure 5e**), will be heavily skewed to lower values by the presence of these fragments. However, quantitative descriptors that are independent of ribbon length would be unaffected by this skew. For example, to calculate the average vertical mass distribution across an entire region, each individual voxel is considered to have the same mass, and the average $z$ location of all voxels relative to the substrate is calculated independent of which voxels are assigned to which ribbon. Similar methods are employed in calculating the average lateral mass distribution, nearest neighbor distance and fraction of interactions, and curvature. The average ribbon length therefore provides a quick means of evaluating the accuracy of the buildup and could be used to gauge the efficacy of future versions of the code.

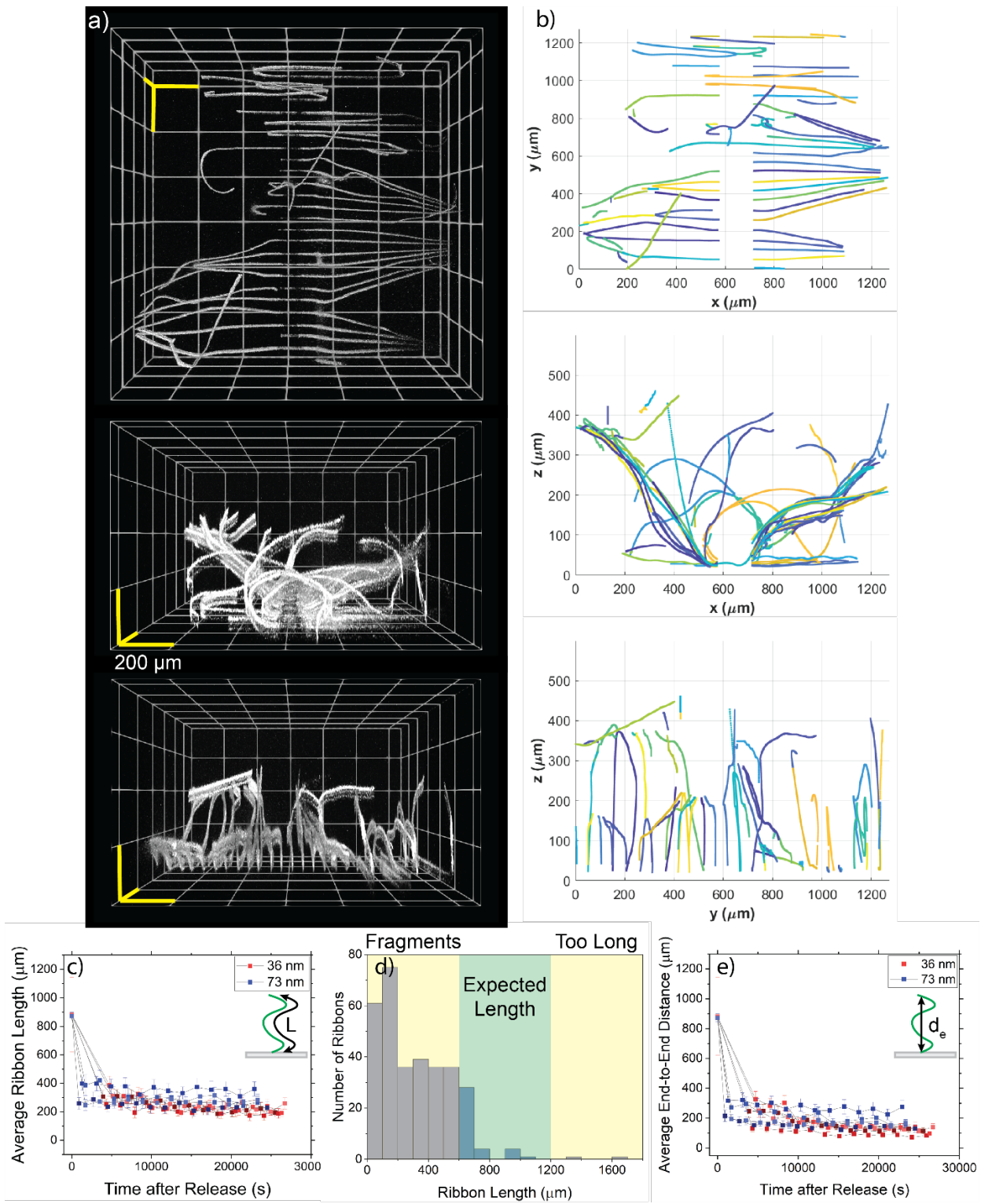

**Figure 5**. Visual and quantitative comparisons of (a) representative confocal microscopy data and (b) its quantitative reconstructions. (c) Ribbon length measurements fall significantly below the expected values as determined by microscopy. (d) The length distribution of the 73 nm thick samples (blue squares in c and e) immediately after release (1300 s – 2100 s) indicates this is due to an overabundance of ribbon fragments. (e) The presence of these fragments

disproportionately affects length-dependent metrics such as ribbon length. We note that in these and future graphs, the red squares of different shades correspond to different regions imaged from the same sample with an average thickness of 36 nm, and the different shades of blue square correspond to different regions of the same sample with an average thickness of 73 nm.

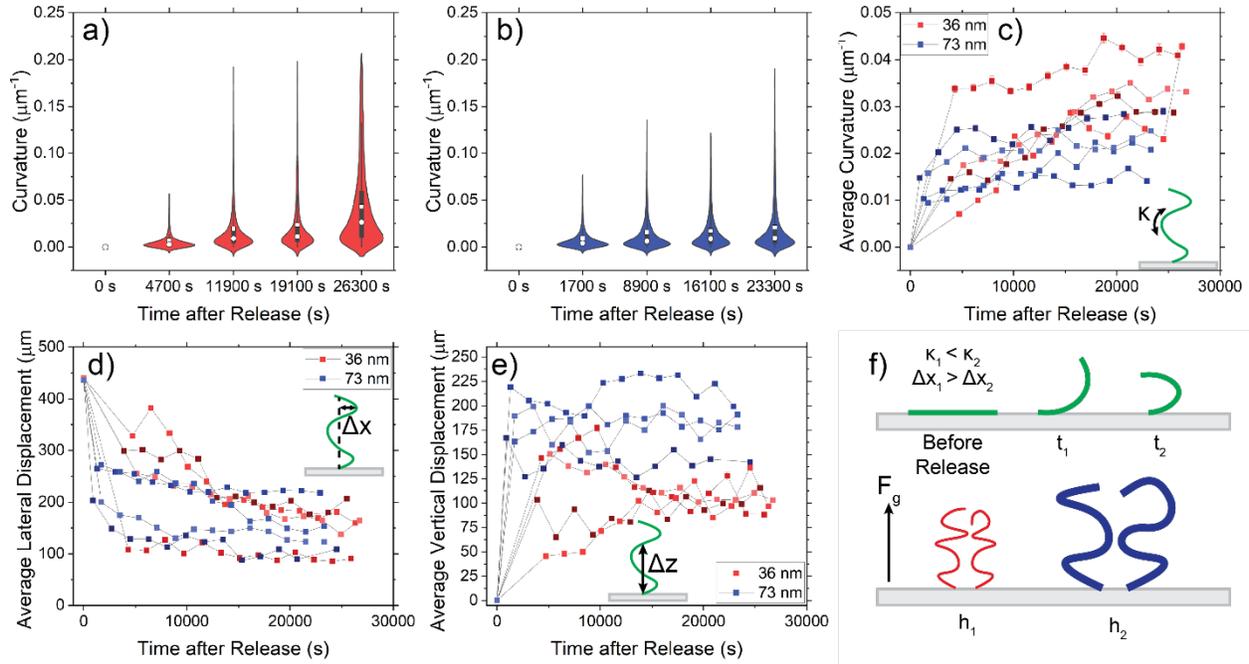

**Figure 6**. The distribution of curvature ($\kappa$) values for (a) 36 nm thick and (b) 73 nm thick ribbons at four different time steps, with quartile values represented along the center axis. Mean and median values are represented by the square and circle, respectively. (c) Average ribbon curvature ($\kappa$) as a function of time after release. Average (d) lateral ($\Delta x$) and (e) vertical ($\Delta z$) displacement across the MSP ribbon arrays as a function of time after release. (f) A visualization of ribbon morphology changes as a function of age and ribbon height. Aging increases ribbon curvature and decreases lateral displacement. Thinner ribbons exhibit higher curvatures and lower vertical displacements. All error bars correspond to a 95% confidence interval.

*Quantitative Comparison of Ribbon Array Architectures*

We compare distribution of 36 nm-thick (**Figure 6a**) and 73 nm-thick (**Figure 6b**) MSP ribbon arrays, which were selected to highlight the impact of thickness on MSP ribbon array morphology with a facile 2:1 scaling. The curvature of the MSP ribbon arrays prior to release is represented at $t$ = 0 s, where the ribbons are flat against the substrate and therefore exhibit zero curvature. After release, both violin plots exhibit an increase in the distribution of curvatures with ribbon aging, with ribbon arrays achieving higher maximum curvatures at later times. The median and mean curvature shift to higher values as the ribbon arrays age. Furthermore, the violin's region of maximum width – corresponding to the most frequent curvature values – broadens over time and similarly shifts to higher curvatures. When considered together, these results indicate that the aging process increases the curvature of MSP ribbon arrays, resulting in higher curvatures at later times, as well as a broader distribution of curvatures across the samples. In **Figure S5d**, we isolate the fraction of each ribbon length where the radius of curvature ($1/\kappa$) is greater than targeted ribbon length of 1000 μm as a means of gauging the straightness of each ribbon. Prior to release, the entirety of a ribbon's length falls within this regime, but following release this fraction is reduced to less than 0.15 of the total length of the ribbon across all samples. This indicates that in all cases, most of the ribbon length is bending after release. These results agree well with the trends of **Figure 6c**, which tracks the average curvature across eight different regions and two different ribbon thicknesses. All results follow a positive correlation between array age and curvature.

**Figure 6d** indicates that the average lateral displacement of the MSP ribbon array decreases with time. The lateral displacement of a point is the distance along the x-axis between a point and its ribbon's anchor to the substrate. The change in lateral displacement as a function of time appears independent of ribbon thickness. Furthermore, when we track the vertical displacement of the MSP ribbon array (**Figure 6e**), we see a distinct thickness correlation but no discernable increases or decreases with time. This indicates that the curvature increase observed in **Figure 6c** likely curls the mesoscale polymer ribbons towards the central line of substrate anchor points, resulting in a decrease in their lateral displacement, but does not greatly impact the average vertical displacement as is depicted schematically in **Figure 6f**. Instead, the average vertical displacement of a ribbon array is strongly influenced by its thickness.

To further explain the origin of this phenomenon, we consider the effects of elastocapillary bending, which favors ribbons that stay closely coiled towards the substrate, and gravity, which tends to extend them away. Given that our ribbon lengths are short relative to their preferred radius of curvature $R_o$, the ribbons exhibit on the order of ~1 or less helical turns. We therefore model them as a circular arc. To estimate the importance of gravity to stretch a filament by unbending some distance $\delta = \Delta z - R_o$, we consider two competing energetic terms. First, we consider the elastic capillary cost to unbend the filament, $U_b = B \left( \frac{1}{\Delta z} - \frac{1}{R_o} \right)^2$, where $B$ is the bending stiffness. Second, we consider the change in gravitational potential energy upon straightening, $U_g = \lambda g \Delta z$, where $\lambda$ corresponds to the mass difference per unit length between the ribbon and the ambient solution. We therefore use the ratio $\delta/R_o$ to parametrize the relative drive of gravity versus the elastocapillary cost of extension, which is very sensitive to ribbon thickness.

For sufficiently thin ribbons, gravitational forces are weak, and ribbons retain their elastocapillary defined shape. Therefore, the negligibility of gravitational effects holds up to a characteristic thickness scale. Given an anticipated surface tension on the order of 70 mN/m,[39] a modulus on the order of 1 GPa,[40,41] and a density differential between polymer and water on the order of $10^{-2}$ g/mL, we estimate a 1 μm characteristic thickness for poly(*tert*-butyl methacrylate) ribbons. Notably, this thickness scale exceeds actual ribbon thicknesses by more than an order of magnitude, implying that gravity effects do not strongly perturb the extension of the ribbons. Therefore, the ribbons' center of mass distribution is set by the preferred, thickness-dependent radius of curvature, i.e. $\Delta z \sim = R_o$ proportional to $h^2$. From **Figure 6e**, we see a mean vertical displacement of 106 ± 28 μm for our 36 nm ribbons and 180 ± 29 μm for our 73 nm ribbons. This 1:1.7 ratio is somewhat smaller than the anticipated 1:4 ratio that our elastocapillary-driven model would predict, and we attribute these differences to the gradient of ribbon thicknesses present in each of the samples.

We track the average distance between nearest neighbor ribbons as a means of gauging interfacial interactions between MSP ribbons. **Figure S5e** depicts no significant correlation for age and the shortest distance between neighboring ribbons. This indicates that there is no change in interfacial interactions between the poly(*tert*-butyl methacrylate) surfaces as the ribbons age in DI water. Given the variance in ribbon widths between 10 μm and 20 μm, we consider an inter-ribbon separation of 10 μm to indicate that two ribbons are close enough for their surfaces to touch and therefore physically and chemically interact. In normalizing the number of these instances relative to the length of the ribbons, we gauge the fraction of each ribbon's length that is interacting with others (**Figure S5f**). Once again, there is no discernable trend in interactions beyond the initial increase that results from the release process and subsequent increase in ribbon mobility. This further supports the finding that aging non-reactive mesoscale polymer ribbon arrays in DI water does not significantly alter their inter-ribbon interactions. We do not observe significant thickness dependence on either metric, and we attribute this to a trade-off between surface area and stiffness. The larger surface area of thicker ribbons should theoretically increase the magnitude of surface interactions between ribbons, however increasing thickness brings with it increased stiffness, and therefore it becomes less energetically favorable to bend to experience repeated interactions with neighbors.

*Mathematical Model of Creep and Ribbon Curvature*

To develop a model that can be used to describe the change in curvature over time and cross-check the computational results, we begin with the Voigt model.[42] This model represents the creep mechanics of a solid material using a spring and dashpot placed in parallel, and approximates the relaxation of the strain $\varepsilon$ with time $t$ as[42]

(2) $\varepsilon(t) = \frac{\sigma_o}{E}(1 - \exp(-t/\tau_o))$ ,

where $\sigma_o$ is the initial, constant, stress, $E$ is the long-time Young's modulus, and $\tau_0 \equiv \eta/E$ is a characteristic relaxation time, which depends on the viscosity $\eta$ of the surrounding fluid. From the literature derivations of helix formation, we conclude that stress acting upon and leading to the deformation of our system arises from the interplay between surface tension and cross-sectional geometry.[24] Assuming that the MSP curvature arises from pure bending, the strain describes elongation and compression of the MSP about a neutral curve of curvature $\kappa$, and thus scales as [43]

(3) $\varepsilon(t) \sim \kappa(t); \kappa \to \frac{\gamma}{Eh^2}$ as $t \to \infty$ .

Therefore, at a fixed distance from the ribbon's central axis, the ribbon's curvature relaxes to its equilibrium value as

(4) $\kappa(t) \sim \frac{\gamma}{Eh^2}\left(1 - \exp\left(-\frac{t}{\tau_o}\right)\right)$ .

In Figure 7, we explore the scaling relationship of two crucial factors in determining the curvature evolution of mesoscale polymer ribbon arrays: thickness $h$ and relaxation time $\tau_o$. As demonstrated in Figure 7a, ribbon thickness $h$ determines the asymptote of the array's curvature, with lower thicknesses yielding a higher curvature following a square scaling relationship. The relaxation time $\tau_o$ determines the time it takes to approach this asymptote, with longer relaxation times resulting in lower initial slopes, and therefore slower approaches.

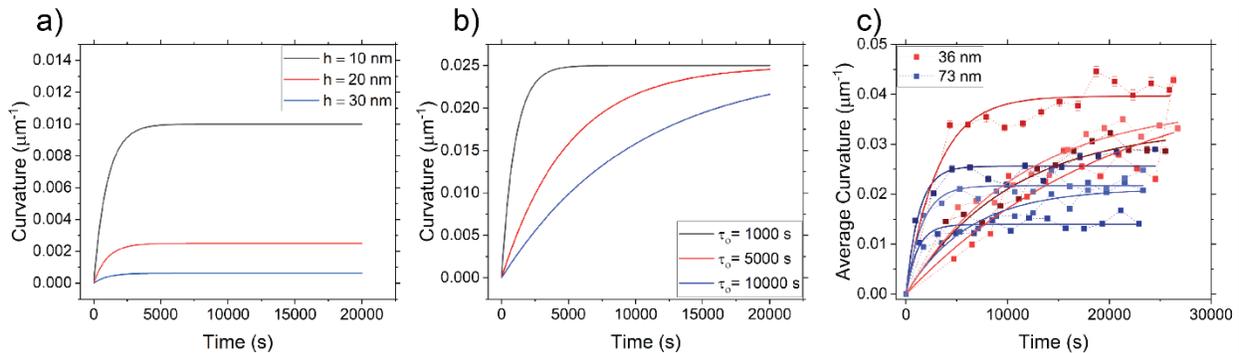

**Figure 7**. Sample scaling of MSP ribbon array curvature as a function of (a) thickness (h) and (b) relaxation time ($\tau_o$). (c) The curvature data of Figure 6c is fitted to Equation 4.

We fit the curvature evolution data from Figure 6c to Equation 8, with results displayed in Figure 7c. The parameters used to calculate the fitted curves are listed in **Table 1**, where we approximate $\gamma/Eh^2$ as curvature $\kappa_o$. The fitted $\kappa_o$ parameters align with Equation 8's correlation between smaller thicknesses and larger curvatures at long time scales. Although the ribbon thicknesses exist at a ratio of 2:1, the ratio between fitted curvature parameters is only 1:1.9 when we would expect it to be in the range of 1:4. In Figure 7c, it appears that several of the 36 nm curves have yet to reach their asymptote, which would result in an underestimation of $\kappa_o$ during fitting.

| Region # | Ribbon Thickness $h$ (nm) | Curvature $\kappa_o$ (μm$^{-1}$) | Relaxation Time $\tau_o$ (s) | Coefficient of Determination $R^2$ | Data Collection Start Time $t_1$ (s) |
|---|---|---|---|---|---|
| 1 | 36 | 0.029 | 5800 | 0.85 | 3900 |
| 2 | 36 | 0.040 | 3110 | 0.92 | 4300 |
| 3 | 36 | 0.045 | 21000 | 0.79 | 4700 |
| 4 | 36 | 0.038 | 11000 | 0.95 | 5100 |
| 5 | 73 | 0.023 | 1300 | 0.91 | 900 |
| 6 | 73 | 0.014 | 1100 | 0.89 | 1300 |
| 7 | 73 | 0.021 | 6200 | 0.90 | 1700 |
| 8 | 73 | 0.022 | 1500 | 0.92 | 2100 |

**Table 1**. Fitting parameters and goodness of fit between average curvature data and creep model.

However, the characteristic time scale $\tau_0$ changes by up to an order of magnitude, though most of the values are on order of $10^3$ s. By the assumption that $\tau_0$ depends only on the material parameters $\eta$ and $E$, we would expect the calculated $\tau_0$ values to converge to a single value. However, we find that the average relaxation time of the 36 nm samples is 10200 ± 7900 s, and the average relaxation time for the 73 nm samples is 2500 ± 2400 s, which results in a 4.1:1 ratio between the two. To explain this disparity, we consider that $\tau_o$ value used by the Voigt model does not account for the geometry of the system. We consider a slender-body mechanical model in which the elastocapillary bending moment is balanced by drag along the ribbon's length, leading to a modified, geometry-dependent relaxation time (see SI for more details),

(5) $\tau_o \sim \dfrac{\eta L^3}{Ewh^3 \ln\left(\frac{L}{w}\right)}$ .

With $\tau_o$ scaling with $1/h^3$, we expect that doubling the thickness of our ribbon arrays would reduce our relaxation time by a factor of 8. These results, coupled with the large variation in fitted values, indicate that our experimental process lacks sensitivity to $\tau_o$, particularly for thicker ribbons. Accurate measurement of $\tau_o$ requires the collection of curvature data immediately following the release of ribbons from the substrate which is extremely difficult to acquire due to the time required in the experimental measurement method. However, the scaling relationship derived in Equation 5 opens ample opportunities to tailor ribbon array relaxation time through material selection, ribbon geometry, and ambient viscosity in future studies.

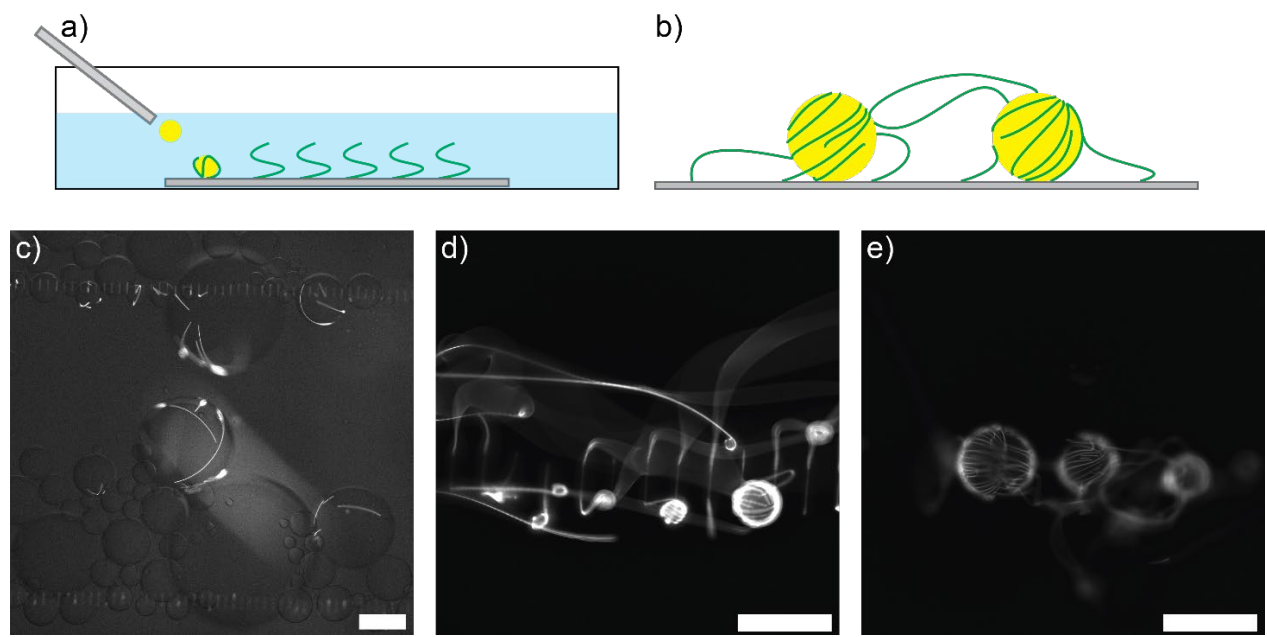

**Figure 8**. (a) Droplets of perfluorodecalin are introduced to an MSP ribbon array release in aqueous solution (b) Mesoscale polymer ribbons wrap the perfluorodecalin droplets. Multiple ribbons wrap each droplet and some ribbons form bridges between the droplets. Optical microscopy images of MSP-droplet interactions in (c) deionized water, (d) pH 10 buffer solution and (e) pH 4 buffer solution indicate this behavior under multiple environmental conditions. All scale bars are 300 µm.

*Towards Applications in Adhesion*

To demonstrate the potential application of mesoscale polymer ribbon arrays as adhesives and filters, we introduced droplets of perfluorodecalin to mesoscale polymer ribbon arrays released into DI water, pH 10, and pH 1 buffer solution (**Figure 8a**). We observe that the mesoscale polymer ribbon arrays spontaneously wrap the oil-water interface, a process previously attributed to capillary forces at the three-phase contact.[14–16] Crucially, as modelled in **Figure 8b** and observed in optical micrographs (**Figure 8c-e**) multiple ribbons spontaneously wrapped individual droplets of perfluorodecalin. Furthermore, **Figure 8e** demonstrates that the mesoscale polymer ribbons are capable of bridging between droplets of perfluorodecalin as well. This experiment provides a preliminary demonstration of how mesoscale polymer ribbon arrays interact with foreign objects in their vicinity, and provides insight with regards to their collective behavior for future applications.

**Conclusion**

In this study, we demonstrate and evaluate a methodology for fabricating and characterizing the morphology of mesoscale polymer ribbon arrays. Our fabrication process enables the creation of long, thin, proximal structures with aspect ratios as high as 100,000. Algorithm-driven image analysis enables the identification of precise ribbon positions in 3D space from confocal microscopy data by determining the most likely ribbon segment connections across junctions. The resulting digital reconstruction can be used to calculate quantitative descriptors of the MSP ribbon arrays such as the radius of curvature, mass distribution, and inter-ribbon separation. The poly(*tert*-butyl methacrylate) ribbons exhibit an increase in curvature as they age through a creep-attributed process, and a mathematical model is proposed to describe this behavior. We observe that the aging process does not pose a significant impact on the surface interactions between ribbons. The methods described here lay the groundwork for a systematic means of probing the impacts of fabrication and environmental conditions on MSP ribbon array morphology. We provide proof-of-concept that arrays of mesoscale polymer ribbons exhibit capillary-driven wrapping behavior reminiscent of the contact splitting behavior found in other high aspect ratio adhesives. This opens up the systems to towards broader impacts in underwater adhesion, as well as microrobotics and biomedical devices.

**Acknowledgements**

This work was supported by the National Science Foundation Graduate Research Fellowship Program (NSF GRFP #1451512) and the University of Massachusetts Amherst Graduate School Spaulding-Smith Fellowship. All confocal microscopy data was gathered at the Light Microscopy Facility at the Institute for Applied Life Sciences at UMass Amherst, and the authors would like to acknowledge the assistance of James Chambers and Maaya Ikeda in developing experimental protocol. All reactive ion etching was performed in the Nanofabrication Cleanroom of the Institute for Applied Life Sciences at the UMass Amherst, and the authors would like to acknowledge the assistance of John Nicholson and Neel Mehta. The authors acknowledge Subhransu Maji, Cornelia Meissner, Xin Hu, and Nolan Miller for helpful discussions, and Dylan Barber for training and guidance during initial exploratory experiments.

**Conflict of Interest**

The authors declare no conflicts of interest.

## Supplementary Information

**Computationally-Quantified Morphology: A Detailed Overview**

*Image Processing.* The microscopy data is deconvolved using NIS-Elements' automatic algorithm and exported to Fiji for processing. We blur the Z-stacks using a 3D gaussian filter then binarize the images via thresholding at value *t*, skeletonize the 3D dataset using Fiji's built-in algorithm, and export the result to a .tiff stack.[28] We manually adjust the binarization threshold *t* for stack of images sample to balance the visibility of the dimmest ribbons with the oversaturation of bright, proximal segments (**Fig 3a-c**).

*Coding Overview.* We use a proprietary MATLAB algorithm to evaluate the morphology of our ribbon arrays. The objective of the code is to first determine the position of individual ribbons in 3D space from the confocal data, and then use the estimated ribbon positions to extract key metrics from the data, such as the radius of curvature, the inter-ribbon separation, and the distribution of mass density relative to the substrate surface. Due to the wide variety of ribbon architectures exhibited across experiments, this strategy allows for a quantitative approach to compare visually distinct morphologies. This code uses strategies from [29–33] to guide its design.

*Skeletonization and Segmentation.* The code takes in the skeletonized .tiff stack and locates the individual (x,y,z) coordinates of each skeleton voxel (**Figure S3a**), with each voxel having dimensions of 1 px x 1 px x 1 *z*-step. Each voxel is classified as an end point, a junction, or a backbone based on the number of pixels in its neighborhood[29] and neighboring backbones are connected to create segments terminated at either end points or junctions (**Figure S3b**)[29,30].

*Bridging Junctions.* The algorithm groups proximal junctions and then identifies the segments that connect to each. To bridge the junctions, we must first describe the direction of each segment's approach. However, because the ribbons can follow relatively tortuous paths, we must first subdivide each ribbon segment into smaller components. In this way, we isolate the region of each segment closest to the junction.

To do this, we define the tortuosity of a ribbon as the difference in the distance between the ribbon's end-to-end distance and its total length. Starting at one end, we trace the segment point-by-point, and once the tortuosity exceeds an empirically-defined threshold, we terminate the growing subsegment and begin a new one.[29] In this way, segments with low curvature are broken into fewer regions of more voxels, and high curvature segments are described by a greater number of subsegments, each containing fewer voxels.

This process leaves us with a list of skeleton subsegments, each described as an *n* x 3 matrix, with each row containing the *x*, *y*, and *z*, coordinates of a voxel. We consider each of these voxels to have equal mass. Using the *cov* function, we compute the gyration tensor of each individual subsegment. The largest principal eigendirection of that gyration tensor is the directionality of the subsegment (**Figure S3c**).

We calculate the principal eigendirection of each subsegment immediately bordering the junction and determine all possible dot products between them (**Figure S3d**). We draw the initial pathways between segments that are the most parallel and should therefore exhibit the largest dot product. The algorithm makes these connections while searching for and ignoring spurious segments– artifacts of the skeletonization process[29,33] – based on the following criteria:

(a) The segment approaches an odd-numbered intersection at an angle perpendicular to all other segments
(b) The segment shares significant (>80%) overlap with another segment in *x* and *y*, with a slight (<10 *z*-step) *z*-offset.

(c) The segment exhibits improbable straightness, with tortuosity $\left(\frac{L_{path}}{L_{e-e}}\right)$ approaching 1.

*Buildup and Evaluation.* Once the initial connections (**Figure 3e**) are made, the algorithm then begins building up connections between segments can be made based on two criteria:

(a) The end points of two segments are within a maximum distance (5 voxels) of one another. In the event that multiple end points meet this criterion, the connection is made to between the segments whose approach is most parallel.
(b) The segments' ends exhibit direct overlap.

The quality of the resulting guess is then evaluated, making the following assumptions about each ribbon:

(a) A ribbon can only have two end points, and therefore cannot branch.
(b) A ribbon is continuous, and therefore cannot have any gaps along its backbone.
(c) A ribbon follows a unique path, and therefore should not have significant overlap with other segments.
(d) Each ribbon is tethered to a substrate at only one end.

If a ribbon fails to pass these criteria, it is broken up based on the criteria that it failed and re-evaluated as separate segments. Ribbons that share >90% of their points with another segment are combined into a single segment. Furthermore, if >20% of a ribbon's points intersect with others, the entire tangle is re-evaluated to determine the most probable path for each ribbon. In this instance, all possible combinations of paths between end points are evaluated to minimize:

(a) The number of points excluded from the buildup
(b) The average number of overlapping points between each of the pathways
(c) The average tortuosity of each proposed pathway
(d) The average difference in path lengths

Buildup and evaluation steps are iterated to generate a final estimate (**Figure 3g**). All segments containing less than 30 points are excluded from consideration. The remaining ribbons are smoothed using MATLAB's *rloess* method due to the method's tolerance of outliers in the data. The window used to smooth these curves is proportional to the size of the segment, and then all ($x,y,z$) coordinates are translated into micron-based positions based on the micron-to-pixel ratio and the size of the $z$-step. $Z$-values are corrected to align the lowest $z$-values with a height of zero.

The algorithm then identifies and removes the portions of ribbons located along the laser cutting line. These ribbons are adhered to the substrate and therefore will not exhibit morphological changes as the environment changes. The algorithm considers all ribbons that are within 20 μm from the slide and less than 200 μm in length, and then it identifies the average location of these ribbons along the $y$-axis. The laser cutting line to which the ribbons are tethered runs perpendicular to the $y$-axis, and therefore the ribbon fragments located along the center line should exhibit a similar range in $y$-values. The algorithm considers the most frequent range in $y$-values to be the location of the center axis. The code then removes all points within 20 μm of the substrate that fall within this $y$-range. This removes the substrate-bound portion of the ribbon from future consideration.

*Quantitative Metrics*.

We quantify the ribbon position using the following metrics:

(a) **Curvature**: determined using the first and second derivatives of the ribbon's coordinates. Assuming the $(x,y,z)$ coordinates of a ribbon trace a path $F(t)$, with a step size of $dt = 1$, then the curvature $\kappa$ can be approximated:

$$\text{(Eq. S1)} \quad \kappa = \frac{\|F'(t) \times F''(t)\|}{\|F'(t)\|^2}$$

(b) **Radius of curvature**: the inverse of the curvature.
(c) **Straightness**: All radius of curvature values greater than the targeted length of the ribbon (1000 µm) are ignored, and the ribbon is considered to be perfectly straight in these regions. The total number of straight points along the ribbon backbone normalized by the number of points along the ribbon length is calculated as a measured of ribbon straightness.
(d) **Lateral Mas Distribution**: The center axis of the laser cutting line is defined by the central line of the excluded region along the laser cut. The average y distance between each ribbon and between the entire set of ribbons relative to this central axis is calculated.
(e) **Vertical Mass Distribution**: each point along the ribbon skeleton is considered to carry the same weight. To determine the height of each ribbon's center-of-mass, the z-coordinates of each ribbon's points are averaged. To determine the vertical distribution of the entire slide, the average of all z-coordinates is evaluated independent of ribbon of origin.
(f) **End-to-end distance**: the Euclidian distance between each ribbon's end points.
(g) **Inter-ribbon separation**: All ribbons other than the ribbon of interest are converted into a point cloud. The distance between each point along the ribbon of interest and its nearest neighbor in the point cloud is determined. The code then averages these distances across each individual ribbon, as well as recorded independently of their associated ribbon
(h) **Number of Inter-Interactions**: As the width of the ribbons can be up to 20 µm, we consider ribbons are close enough to interact with one another if they are within 10 µm of each other. We identify and count the number of instances where the inter-ribbon separation is less than or equal to this cutoff.
(i) **Length**: Sum of the distance between each pair of neighboring points along the ribbon backbone.

**Derivation of Curvature and Relaxation Time**

The total energy $U$ of a ribbon, as a function of time $t$, is given by:

$$\text{(Eq. S2)} \quad U(t) = \int_0^L ds \left\{ \frac{1}{2} E I_{yy} \kappa^2(s,t) - (\gamma P \Delta X_y) \kappa(s,t) \right\}$$

Where $E$ is the Young's modulus, $I_{yy}$ is the second moment of area of the ribbon's cross-section, $\kappa(s,t)$ is the curvature at position $s$ and time $t$, $P$ is the cross-sectional perimeter, and $\Delta X_y$ is the offset between the centers of the ribbon's cross-sectional perimeter and cross-sectional area. Assuming that the ribbon lies along the $x$-axis with its thickness along the $y$-axis, its moment about the $z$-axis, $M_z(s,t)$ is described as

$$\text{(Eq. S3)} \quad M_z(s,t) = E I_{yy} \kappa(s,t) - \gamma P \Delta X_y$$

We assume that the ribbon deforms only in the $xy$-plane, and is therefore described by the reduced Kirchhoff rod equilibrium equations, namely[44]

$$\text{(Eq. S4)} \quad \frac{d\boldsymbol{F}}{ds} = -\boldsymbol{f}(s,t), \text{ and}$$

$$\text{(Eq. S5)} \quad \frac{dM_z}{ds} = (\boldsymbol{F}(s,t) \times \hat{\boldsymbol{t}}(s,t))_z \; .$$

Here, $\hat{\boldsymbol{t}}(s,t)$ is the unit tangent vector along the ribbon, and $\boldsymbol{F}(s,t)$ is the internal force along the ribbon and $\boldsymbol{f}(s,t)$ is the external force density along the ribbon.

We next assume that each segment of the ribbon is subject to Stokes drag. Therefore, each point $\boldsymbol{r}(s,t)$ along the ribbon will experience a force proportional to the velocity $\boldsymbol{v}(s,t) = \partial_t \boldsymbol{r}(s,t)$. The force per unit length is given by

$$\text{(Eq. S6)} \quad \boldsymbol{f}(s,t) = -\zeta \frac{d\boldsymbol{v}}{ds} ,$$

where $\zeta$ is a drag coefficient. By integrating over $s$, the internal force becomes

$$\text{(Eq. S7)} \quad \boldsymbol{F}(s,t) - \boldsymbol{F}(0,t) = \zeta \int_0^s ds' \frac{d\boldsymbol{v}}{ds'} = \zeta (\boldsymbol{v}(s,t) - \boldsymbol{v}(0,t)) \; .$$

Finally, we assume that the ribbon is attached to the substrate at $s = 0$, leading to the boundary conditions $\boldsymbol{v}(0,t) = 0$ and $\boldsymbol{F}(s,t) = 0$, and thus

$$\text{(Eq. S8)} \quad \boldsymbol{F}(s,t) = \zeta \boldsymbol{v}(s,t) \; .$$

When considering the moment balance, we express the velocity at segment $s$ in terms of the local tangent and normal frame $\hat{\boldsymbol{n}}$ as

$$\text{(Eq. S9)} \quad \boldsymbol{v}(s,t) = v_t(s,t) \hat{\boldsymbol{t}}(s,t) + v_n(s,t) \hat{\boldsymbol{n}}(s,t)$$

Noting that the tangential component $v_t$ does not contribute to the moment $M_z$ and $\hat{\boldsymbol{t}} \times \hat{\boldsymbol{n}} = \hat{\boldsymbol{z}}$, we simplify the mechanical equilibrium condition to

$$\text{(Eq. S10)} \quad \frac{dM_z}{ds} = -\zeta v_n(s,t) \; .$$

We next assume that the ribbon is constrained to lie flat against the substrate for $s \leq 0$, which corresponds to the position of the ribbon anchor. The ribbon curves along the interval $0 < s < L$, so integrating Eq. S10 from $s = 0$ to $s = L$, we find that

$$\text{(Eq. S11)} \quad M_z(L,t) = -\zeta \int_0^L ds\, v_n(s,t) = -\zeta L \langle v_n \rangle(t) ,$$

where the normal component of the velocity averaged along the length of the ribbon is represented as

$$\text{(Eq. S12)} \quad \langle v_n \rangle(t) = \frac{1}{L} \int_0^L \mathrm{d}s\, v_n(s,t) \ .$$

We consider a minimal model of ribbon shape that satisfies the shape constraints, namely

$$\text{(Eq. S13)} \quad \boldsymbol{r}(x,t) = \hat{\boldsymbol{x}} x + \hat{\boldsymbol{y}} \frac{k(t)}{2} x^2 \ ,$$

where positions along the ribbon are expressed in terms of the $x$-coordinate. To leading order in the small deflection approximation, $s \approx x$ and furthermore

$$\text{(Eq. S14)} \quad \hat{\boldsymbol{t}}(x,t) = \frac{\mathrm{d}\boldsymbol{r}}{\mathrm{d}s} \approx \hat{\boldsymbol{x}} + \hat{\boldsymbol{y}} k(t) x \ ,$$

$$\text{(Eq. S15)} \quad \hat{\boldsymbol{n}}(x,t) = \hat{\boldsymbol{z}} \times \hat{\boldsymbol{t}}(x,t) \approx \hat{\boldsymbol{y}} - \hat{\boldsymbol{x}} k(t) x \ , \text{ and}$$

$$\text{(Eq. S16)} \quad \kappa(x,t) = \frac{\mathrm{d}^2 \boldsymbol{r}}{\mathrm{d}s^2} \cdot \hat{\boldsymbol{n}}(x,t) \approx k(t) \ .$$

The normal component of the velocity and its average are therefore

$$\text{(Eq. S17)} \quad v_n(x,t) = \frac{1}{2} \dot{k} x^2 \text{ and}$$

$$\text{(Eq. S18)} \quad \langle v_n \rangle(t) \approx \frac{1}{L} \int_L^0 \mathrm{d}x\, v_n(x,t) \approx \frac{1}{6} \dot{k} L^2 \ .$$

This yields the moment balance:

$$\text{(Eq. S19)} \quad \dot{k} + \frac{6 E I_{yy}}{\zeta L^2} k(t) = \frac{6 \gamma P \Delta X_y}{\zeta L^2} \ .$$

We integrate and assume that $k(0) = 0$ to yield

$$\text{(Eq. S20)} \quad k(t) = \kappa_o \left( 1 - e^{-\frac{t}{\tau_o}} \right),$$

where

$$\text{(Eq. S21)} \quad \kappa_o = \frac{\gamma P \Delta X_y}{E I_{yy}} \text{ and}$$

$$\text{(Eq. S22)} \quad \tau_o = \zeta L^2 / 6 E I_{yy}$$

Given that $I_{yy} \sim w h^3$ and that the drag coefficient of long, thin rods can be approximated as[45]

$$\text{(Eq. S23)} \quad \zeta \approx \frac{4\pi \eta L}{\ln\left(\frac{L}{w}\right)}$$

we can therefore solve for the relaxation time:

$$\text{(Eq. S24)} \quad \tau_o \sim \frac{\eta}{E} \frac{L^3}{w h^3 \ln\left(\frac{L}{w}\right)} \ .$$

**Figures**

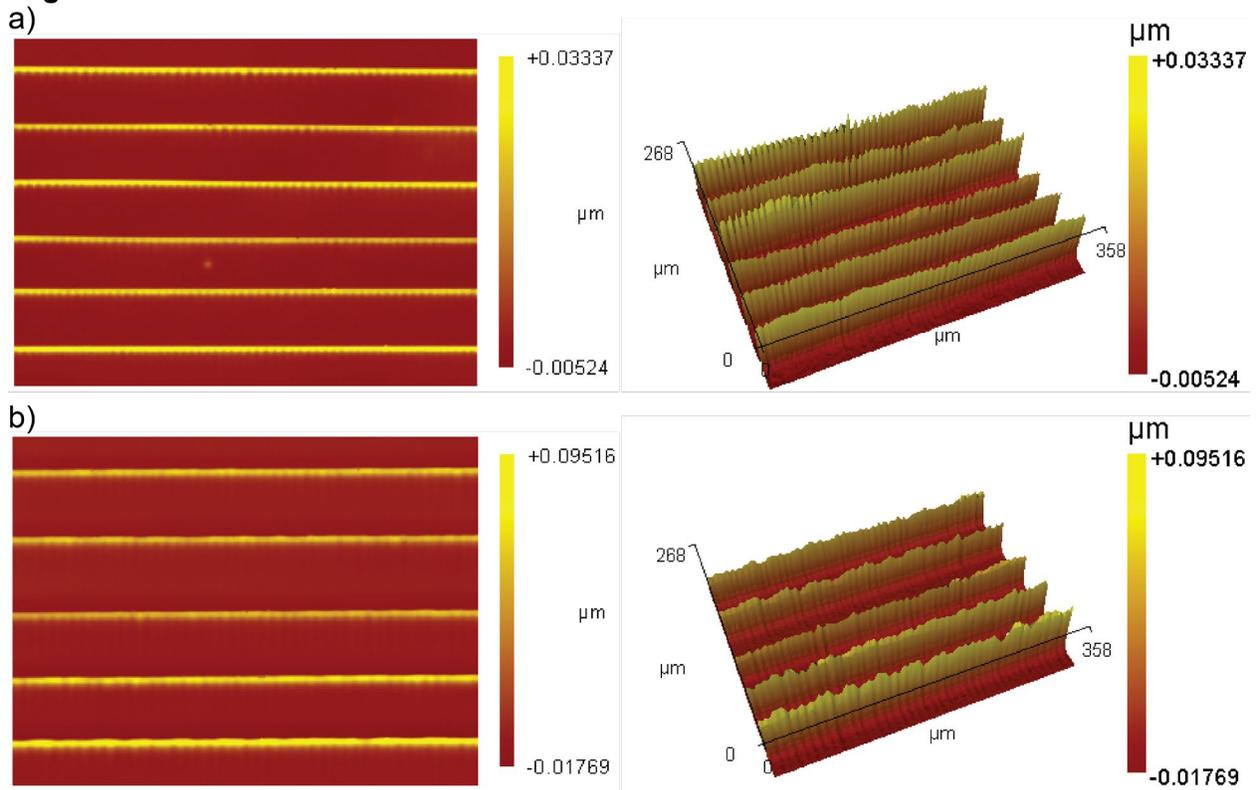

**Figure S1**. Optical profilometry profiles of two different samples, measured to have an average thickness of (a) 36 nm ± 7nm and (b) 73 nm ± 40 nm across the slide, respectively.

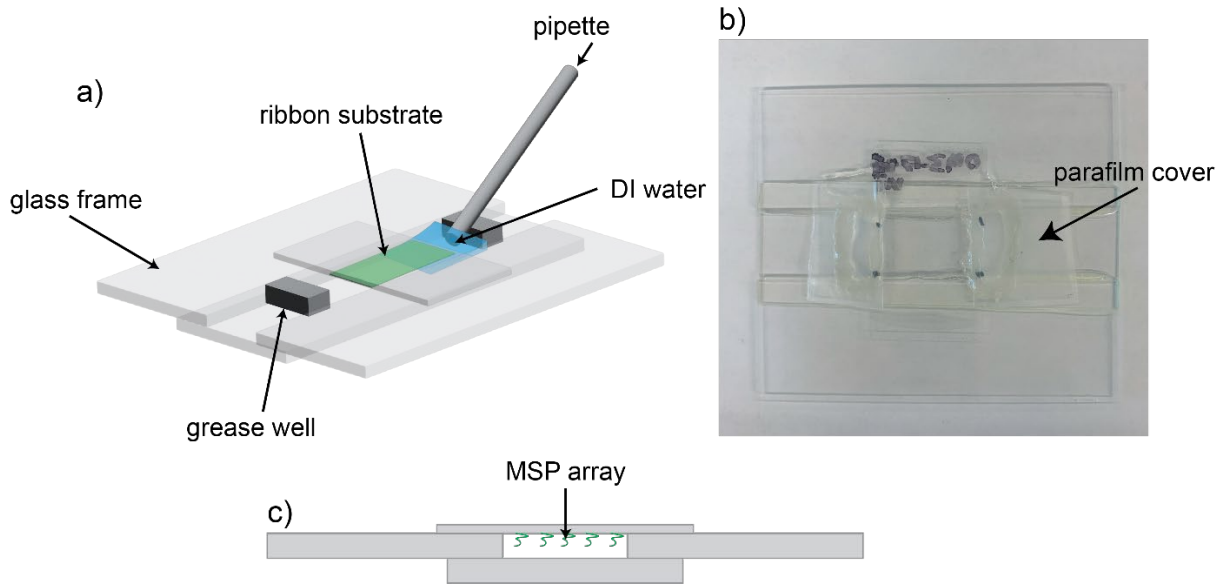

**Figure S2**. Model of a unidirectional flow chamber used for confocal microscopy. The MSP ribbon array hangs from the cover slip down into the flow chamber, as can be seen in (c).

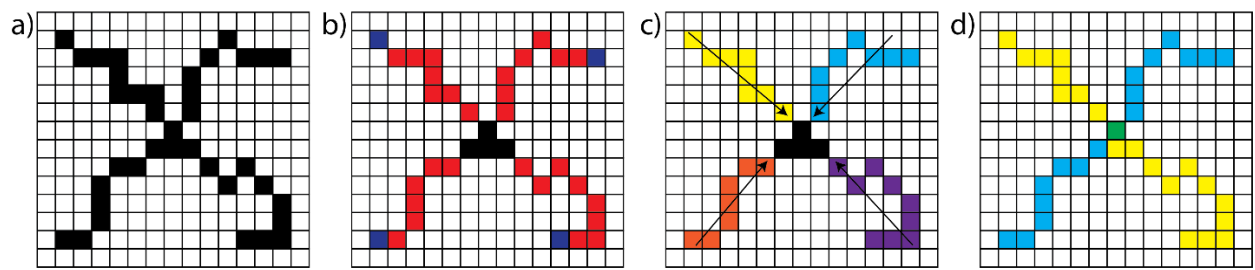

**Figure S3**. A 2D representation of the process of building up ribbons in 3D. (a) Individual (x,y) pixel values are sorted (b) into end points (blue), backbones (red), and junctions (black). (c) Neighboring pixels and backbones are connected, and their directionality relative to their neighboring junction is identified. (d) The final buildup of two ribbons (yellow, blue) and their overlapping sections (green).

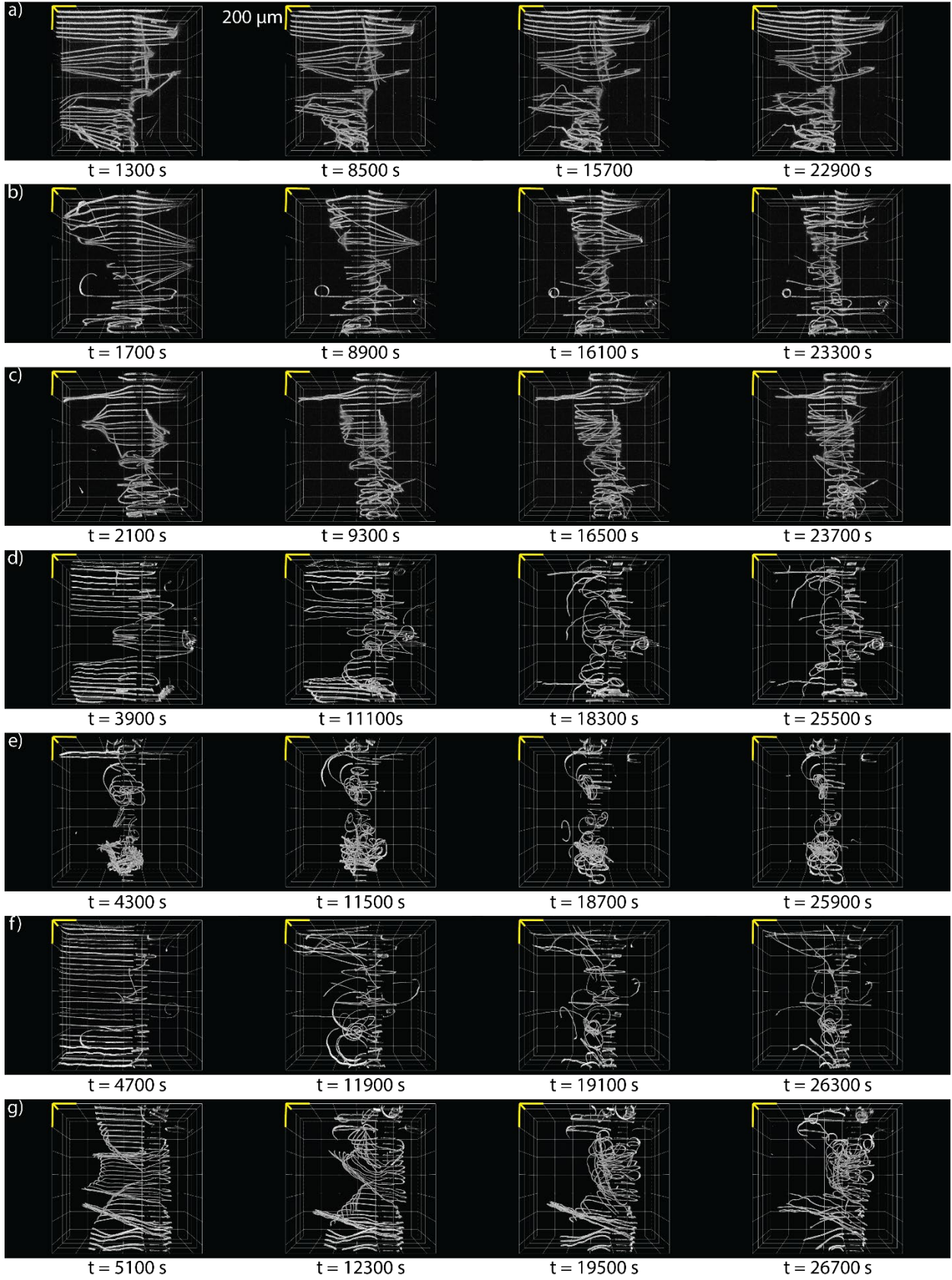

**Figure S4**. Morphological changes of (a-c) 72 nm and (d-g) 36 nm thick MSP ribbon arrays. All scale bars represent 200 µm.

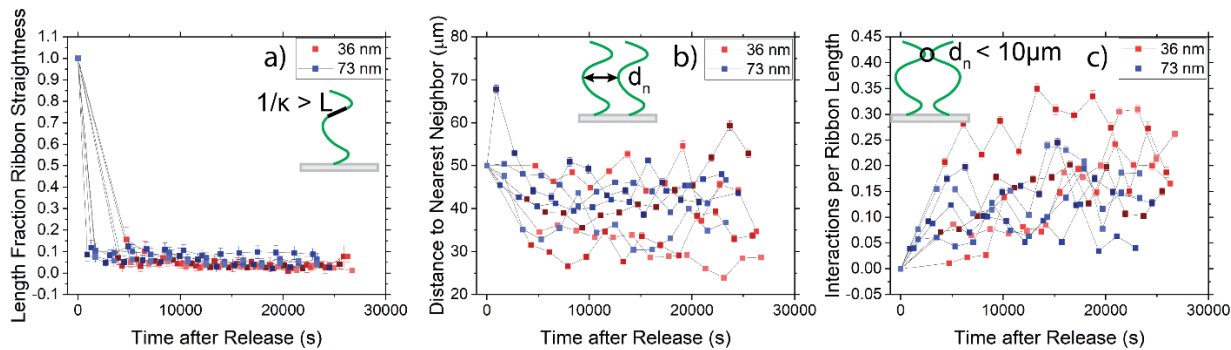

**Figure S5**. Additional computationally-determined quantitative descriptors. (a) The fraction of each ribbon length wherein the radius of curvature (1/κ) is greater than the theoretical ribbon length, suggesting that the ribbon is perfectly straight in these regions. (b) The average nearest neighbor distance between ribbons. (c) The fraction of instances where the distance between nearest neighbor ribbons are less than the width of a ribbon, suggesting that the surfaces are close enough to interact. All error bars correspond to a 95% confidence interval.